# Thermodynamic Properties of the Kagomé Lattice in Volborthite

Satoshi YAMASHITA[+], Tomoya MORIURA, Yasuhiro NAKAZAWA*, Hiroyuki YOSHIDA[1], Yoshihiko OKAMOTO[2], and Zenji HIROI[2]

*Department of Chemistry, Graduate School of Science, Osaka University, Machikaneyama 1-1, Toyonaka, Osaka 560-0043, Japan*

[1]*Superconducting Materials Center, National Institute for Materials Science (NIMS), 1-1 Namiki, Tsukuba, Ibaraki 305-0044, Japan*

[2]*Institute for Solid State Physics, University of Tokyo, 5-1-5, Kashiwanoha, Kashiwa, Chiba 277-8581, Japan*

Thermodynamic investigations on volborthite ($Cu_3V_2O_7(OH)_2 \cdot 2H_2O$), which possesses a two-dimensional kagomé structure of $S$-1/2 spins, are presented. The low-temperature heat capacities of volborthite and its Zn analogue compound were measured by the relaxation calorimetry technique between 0.8 and 45 K. The magnetic heat capacity of volborthite is characterized by two contributions of $T$-linear and $T^2$ dependences, the former of which is large above 1 K, giving evidence of dense gapless excitations. We find a distinct kink in $C_p T^{-1}$ at $T^* = 1$ K, which demonstrates a thermodynamic phase transition of short-range nature to a novel ground state as reported in previous $^{51}$V-NMR experiments. The $T$-linear term becomes smaller but remains finite even in the low-temperature state below $T^*$, while it is gradually reduced with increasing magnetic field and vanishes at 5 T, which is close to the critical field for the field-induced transition observed in the first magnetization step.

*E-mail address: nakazawa@chem.sci.osaka-u.ac.jp

[+] Present address: *RIKEN, 2-1, Hirosawa, Wako, Saitama 351-0198, Japan*

The peculiar low-temperature magnetic properties of frustrated spin systems with antiferromagnetic interactions have been a fascinating topic in condensed matter science over the past three decades.[1-3] The two-dimensional (2D) triangular and kagomé lattices or three-dimensional spinel and pyrochlore lattices *etc.* are realistic stages where frustration physics can be pursued.[4] In these triangle-based structures, geometrical frustration prevents the formation of long-range orderings even if the nearest-neighbor magnetic interaction $J/k_B$ is strong enough. The so-called Néel-type order does not exist, especially, in the case of $S$-1/2 systems with a strong quantum character. Theories have predicted novel possibilities for the ground states: the long-range RVB model, which is expressed by resonating singlet bonds,[5] and other spin liquids with small-gap or gapless characters are discussed.[1]

From the experimental side, however, suitable candidate materials for exploring frustration physics are limited, because residual disorders in a crystal and coupling with other degrees of freedom such as lattices and charges tend to stabilize ordered or glassy states rather than quantum liquid states.[6-7] A quantum spin liquid state for the $S$-1/2 triangular system has recently been realized in organic dimer-based complexes of $\kappa$-(BEDT-TTF)$_2$Cu$_2$(CN)$_3$[8] and EtMe$_3$Sb[Pd(dmit)$_2$]$_2$,[9] which are known as 2D Mott insulators. In these compounds, owing to their ideal two-dimensionality and strong quantum fluctuations enhanced by strong correlation effects, long-range ordering does not occur down to 30 mK and a gapless ground state has been suggested,[8,10] as evidenced by the observation of a $T$-linear contribution in heat capacity. In contrast, the appearance of a tiny gap as small as $\Delta = 0.0016J$ was reported in thermal conductivity measurements.[11]

The 2D $S$-1/2 kagomé lattice with antiferromagnetic interactions is considered as a more frustrated system than a triangular lattice, because the 120º structure that is possible in the 2D triangular Heisenberg model is not stabilized in this system.[4] Therefore, kagomé antiferromagnets are more likely to exhibit a spin-liquid state as a ground state. The issue whether or not a gap exists in low-energy excitations remains unresolved both theoretically and experimentally.[12-13] Recently, natural mineral compounds such as volborthite,[14] herbertsmithite,[15] and vesignieite[16] have been recognized to be model materials of the 2D kagomé antiferromagnet. In vorborthite with a chemical formula of Cu$_3$V$_2$O$_7$(OH)$_2 \cdot$2H$_2$O, the basic plane is composed of CuO$_4$ square plaquettes and a divalent copper ion with $S$-1/2 forms a kagomé spin lattice. The 2D sheets are well separated by a pillared structure of vanadium oxides and H$_2$O molecules. Since the constitutive unit is an isosceles triangle, the kagomé structure is not perfect but should contain two magnetic interactions, $J/k_B$ and $J'/k_B$, as reported in ref. 14. The distortion ratio defined by the parameter $\alpha = J'/J$ is estimated to be 1.2.[17] The average of magnetic interactions, $J_{av}/k_B$, is defined as $(2J/k_B + J'/k_B)/3$. Magnetic order was not observed down to 1.8 K in the first study by Hiroi *et al.*[14] but a glasslike behavior was suggested in a subsequent work.[18] Recently, H. Yoshida *et al.* have succeeded in markedly improving the sample quality by adopting the hydrothermal annealing method.[19] Using a well-annealed sample with an impurity level less than 0.1%, they have shown that the glassy state is not intrinsic but is possibly due to impurity spins, and that the low-temperature spin state is a gapless or tiny gapped ($\Delta < J/1500$) spin-liquid state. Furthermore, they have reported anomalies at $T^* = 1$ K determined by $^{51}$V-NMR experiments. The following NMR study by M. Yoshida *et al.* [20] revealed that certain magnetic order sets in below $T^*$. However, the low-temperature phase shows anomalous properties such as a Lorenzian line shape, $T$-linear behavior in the nuclear spin-lattice relaxation rate $1/T_1$ indicating dense low-energy excitations, and a large spin-echo decay rate $1/T_2$ pointing to unusually slow fluctuations. The compound also exhibits three magnetization steps under high magnetic fields of 4.3, 25.5, and 46 T.[19] The NMR experiments found a magnetic-field-induced phase transition in the first step field.[20]

To discuss the low-energy problems of frustrated spins especially in terms of entropy, thermodynamic studies using well-characterized samples are necessary. Here, we report the results of systematic heat capacity measurements using a volborthite sample with improved quality. The possibility of the formation of spin-liquid states and the nature of the 1 K anomaly are investigated thermodynamically.

Powder samples of volborthite $Cu_3V_2O_7(OH)_2 \cdot 2H_2O$ and its Zn analogue compound $Zn_3V_2O_7(OH)_2 \cdot 2H_2O$ were synthesized by the precipitation method described previously.[14] To improve the sample quality, hydrothermal annealing at 190 ºC for 12 h was performed. This annealing process improved the crystallinity of the Cu compound, as evidenced by X-ray diffraction measurements.[19] In contrast, it was less effective for the Zn compound, so that its sample quality was poor in comparison with the Cu compound. Heat capacity measurements were performed by the thermal relaxation method, which is appropriate for measuring the absolute value of $C_p$ even for small amount of sample less than 1 mg. We made a small pellet of typically about 0.2-0.4 mm thickness, 1.5 mm diameter, and 0.7-1.4 mg weight. We used two different calorimetry cells mounted on $^3$He and $^4$He cryostats, both of which were designed by ourselves.[21-22] The temperature ranges covered by these apparatus are 0.8 -12 K for the $^3$He system and 9 - 45 K for the $^4$He system. Although the data were obtained with two different calorimeters using different pellets, the data coincided with each other in the overlapped temperature region between 9 and 12 K.

In Fig. 1, the temperature dependences of the heat capacities of Cu volborthite and the Zn analogue are shown in a $C_p$ vs $T$ plot. The absolute values of the heat capacity of the Cu compound are for example $4.97 \times 10^{-2}$ J K$^{-1}$ Cu-mol$^{-1}$ at 1 K, $1.19 \times 10^0$ J K$^{-1}$ Cu-mol$^{-1}$ at 10 K, and $1.56 \times 10^1$ J K$^{-1}$ Cu-mol$^{-1}$ at 40 K in the unit per 1 mol of Cu ions. Since the heat capacity increases rapidly with temperature, the $C_p$ data are shown on a logarithmic scale to see the overall behavior in the whole temperature range. The data at low temperatures below 13 K are also shown in a $C_p T^{-1}$ vs $T$ plot in Fig. 2 together with the data obtained under magnetic fields for the Cu volborthite compound.

Reflecting frustration effects in the kagomé lattice, the compound does not show any peak indicative of long-range magnetic order down to 0.8 K, except for a small kink, which will be discussed later. Instead, it shows a large heat capacity compared with the Zn analogue at low temperatures, which demonstrates that the magnetic entropy of $S$-1/2 survives down to a low-temperature region. These features are consistent with the heat capacity previously reported by Hiroi *et al.* for an initial-stage sample that was not subjected to a hydrothermal annealing process.[14] However, the absolute values of heat capacity and its temperature dependence especially at low temperatures are substantially different between the present and previous data. Since the previous sample contained a large amount of impurity spins that cause spin glass behavior, the heat capacity reported in ref. 14 might have been masked by this glasslike behavior, which gives a relatively large heat capacity at low temperatures. It is likely that intrinsic properties associated with the spin frustration have appeared in the present data in Fig. 1 for a much cleaner sample.

In order to discuss the magnetic heat capacity of the Cu compound, the estimation of the lattice heat capacity is necessary. In general, the substitution of a nonmagnetic ion for a magnetic one in an isostructural crystal is an effective method of deriving a common lattice contribution. The temperature dependence of the Debye temperature [23] of the Zn compound shown in Fig. 1(b) is complicated and is different from the simple Debye model in the low-energy region, which is probably due to the existence of soft phonon modes originating from $H_2O$ molecules in the interlayer space, as in molecular compounds. In addition to this, the Zn compound has a slightly different crystal structure owing to the lack of the Jahn-Teller distortion.[24] This may be the reason why the heat capacity of the Zn compound is slightly larger than that of the Cu compound above 25 K. Therefore, accurate estimation of the lattice heat capacity from the data of Zn compound is difficult.

Taking these inconvenience factors of the Zn compound as reference into account, we have tentatively determined a lattice contribution for volborthite according to the following procedure. At first, we subtracted the residual $T$-linear contribution with a coefficient of 2.6 mJ $K^{-2}$ Zn-$mol^{-1}$ from the heat capacity data of the Zn compound, assuming that it originated from a small amount of parasitic impurity phase. Note that the sample quality of the Zn compound is poor as compared with that of the Cu compound. Second, we multiplied the data by a constant scale factor of 0.89 to adjust the heat capacity of the Zn compound to that of the Cu compound at 45 K. The resultant lattice contribution is shown in Figs.1 and 2.

By subtracting this lattice contribution from the data of the Cu compound, the magnetic heat capacity is obtained, as shown in Fig. 3(a). It has a broad hump with a maximum at approximately 8 K. This temperature corresponds to 0.1 of the $J_{av} / k_B$ value of 77 K according to preliminary data by Hiroi *et al.* The theoretical work by Elser[25] and the more recent work by Wang *et al.*[26] have predicted that the heat capacity has a broad hump at around $0.7 J / k_B$ and a low-temperature peak at around $0.1 J / k_B$. Wang *et al.* calculated the effect of distortion from the ideal kagomé lattice using the parameter $\alpha = J'/J$ and confirmed that the peak structure is retained even at 25% of distortion. The peak observed at ~8 K may correspond to this low-temperature peak, though the peak height is much smaller that expected. By integrating $C_{mag}T^{-1}$ with respect to temperature, the temperature dependence of magnetic entropy from $Cu^{2+}$ spins is derived as shown in Fig. 3(b). Compared with the expected total entropy, $Rln2 = 5.76$ JK$^{-2}$mol$^{-1}$, the calculated magnetic entropies are 11.9% at 10 K and 17.7% at 15 K. Elstner and Young have found a missing entropy of 40% at $T=0$ in an $S$-1/2 kagomé system.[27] The observed smaller experimental entropy for volborthite may be related to this missing entropy. However, we have to be careful in concluding this, since the magnetic entropy should exist above 45 K even if the major contribution is given at approximately 20 K where the magnetic susceptibility shows a broad peak.[14] The lattice curves shown in Figs. 1 and 2 probably overestimate the actual lattice contribution, especially in the higher temperature range. Further investigation is required to evaluate the entropy, particularly at high temperatures.

Figure 4 shows a low-temperature part of the $C_{mag}T^{-1}$ vs $T$ plot. Three characteristic features are to be noted in the 0 T data. The first one is that the $C_{mag}T^{-1}$ shows a linear temperature dependence between 1.1 and 3.5 K, that is, a $T^2$ term with a coefficient of 8.29 mJ K$^{-3}$ Cu-mol$^{-1}$ exists in the magnetic heat capacity. Second, a linear extrapolation of the $C_{mag}T^{-1}$ in this temperature range to $T = 0$ gives a finite value of 40.1 mJ K$^{-2}$ Cu-mol$^{-1}$. This finite value corresponds to the so-called γ term usually defined as an electronic heat capacity coefficient in metallic compounds. Note that volborthite is highly insulating without conduction electrons. Such a linear contribution of a magnetic insulator is expected for an antiferromagnet with a one-dimensional character, which may not be the present case. Thus, the existence of the $T$-linear contribution suggests that certain gapless excitations in spin-liquid state are realized in this temperature region. In triangular compounds such as κ-(BEDT-TTF)$_2$Cu$_2$(CN)$_3$, the quantum spin liquid behavior has been evidenced by the observation of a similar $T$-linear term.[10] The corresponding value of the present kagomé compound is almost twice as large as those of the triangular compounds, demonstrating that the density of states for spin excitations is larger in the kagomé compound. The third characteristic feature to be noted in Fig. 4 is the existence of a kink at 1 K, which is the same temperature ($T^*$) at which a sharp peak was observed in the spin-lattice relaxation rate in previous NMR experiments. The previous heat capacity data reported by H. Yoshida *et al.* in ref. 19, which was measured using a commercial calorimeter (Quantum-Design PPMS system), show a broad hump at around 1 K, possibly corresponding to this kink structure. The present measurements have unveiled the structure clearly, because we measured on the better-quality sample in a better experimental configuration ensuring a good thermal contact between the thermometer and the sample. The kink suggests a thermodynamic phase transition with a short-range character or a kind of freezing in spin states at a low energy. Below 1 K, the $C_{mag}T^{-1}$ data seem to obey a different linear relation against $T$, as shown in Fig. 4; the coefficient of the $T^2$ term is increased to 19.9 mJ K$^{-3}$ Cu-mol$^{-1}$ and the residual $T$-linear term, $\gamma^*$, evaluated as a section of the vertical axis is reduced to 15.2 mJ K$^{-2}$ Cu-mol$^{-1}$. Therefore, the low-temperature phase is also considered to be gapless with a low density of states.

The low-temperature heat capacities at magnetic fields of 1, 3, 5, and 7 T are also shown in Figs. 3(a) and 4. The temperature dependence of $C_{mag}T^{-1}$ and the kink structure observed at 0 T do not change at $H = 1$ T. In contrast, the higher magnetic fields change the heat capacity systematically: as $H$ increases, the amount of $T$-linear component decreases gradually and the kink tends to disappear. At the same time, the broad peak at around 8 K at $H = 0$ T grows by about 20% at 7 T with the peak temperature slightly shifted to lower temperatures, as shown in Fig. 3(a).

Hereafter, we discuss these thermodynamic data with reference to other experiments to disclose the nature of the low-temperature phase of volborthite. The magnetic susceptibility of volborthite has been studied down to 60 mK, which shows a large residual $\chi(0)$ of $3.0\times10^{-3}$ cm$^3$ Cu-mol$^{-1}$ at 1 T and no anomaly at 1 K.[19] Using this $\chi(0)$, one may evaluate the ratio between $\chi(0)$ and the magnitude of the $T$-linear component in heat capacity, which is known as the Wilson ratio $R_W$ for a metallic system: $R_W = 1$ for a free electron system and 2 for a correlated electron system. These Wilson ratios for volborthite are unusually large : 5.5 for the high-temperature phase above 1 K and 15.4 for the low-temperature phase below 1 K. In contrast, the organic $S$-1/2 triangular system of κ-(BEDT-TTF)$_2$Cu$_2$(CN)$_3$ gives a Wilson ratio close to unity.[10] The rather large $R_W$ values in the present system may indicate an unusual gapless state with dense low-energy excitations, which is substantially different from that observed in the Fermi liquid.

The appearance of the low-temperature phase below 1 K in the kagomé system is interesting from the thermodynamic viewpoint. Although heat capacity does not show any divergent peak signaling the formation of a long-range ordering, the clear kink structure across this temperature demonstrates that a certain type of phase transition takes place as a bulk property. However, the magnetic entropy associated with this phase transition is only less than 1% of $Rln2$. This small entropy and absence of clear anomaly in the magnetic susceptibility, as reported by H. Yoshida *et al.*,[19] indicate that the phase transition is attributed to a kind of short-range order produced by the residual degree of freedoms in frustrated spins. It is noted that the low-temperature phase is also gapless and low-energy excitations still exist with spins fluctuating owing to the frustration effects. Note that the $^{51}$V-NMR experiments by M. Yoshida *et al.*[20] have shown that the low-temperature phase has an anomalous feature characterized by very slow dynamics. Although the origin of this interesting low-temperature phase is not clear at present, it must involve a new physical aspect related to the many-body effects in the kagomé system.

From the low-temperature heat capacity data under magnetic fields shown in Fig. 4, it is emphasized that this low-temperature phase is stable upon applying weak magnetic fields: the data at 1 T coincide well with those at 0 T. Under a magnetic field of 3 T, the kink temperature $T^*$ seems to shift to a higher temperature and $\gamma^*$ is reduced down to about 9.6 mJ K$^{-2}$mol$^{-1}$. With further increase in magnetic field, $\gamma^*$ decreases and finally vanishes at 5 and 7 T, which demonstrates that the ground state has transformed to a different state above 5 T. This critical field is nearly equal to $H_{s1} = 4.3$ T for the magnetic-field-induced transition observed in the first magnetization step. Therefore, we conclude that the low-energy excitations existing in the low-field phase are completely removed in the high-magnetic-field phase. Moreover, the transition between them occurs continuously with $\gamma^*$ decreasing gradually with increasing magnetic field. The growth of the $C_{mag}T^{-1}$ peak at 8 K with an increase in magnetic field, as shown in Fig. 3(a) means that the magnetic entropy at low temperatures shifts to the higher temperature side in the presence of a magnetic field. These thermodynamic features observed in this low-energy region may give an interesting issue related to an order parameter for the "hidden" phase found in volborthite.

Finally, we summarize the thermodynamic properties of volborthite revealed using the hydrothermally annealed sample. We have observed a clear kink at 1 K at zero magnetic field in $C_pT^{-1}$, which evidences a thermodynamic phase transition of higher order or a crossover. Low-energy excitations are featured by the presence of $T^2$- and $T$-linear terms in heat capacity, the coefficients of which change dramatically across the transition: 8.29 mJ K$^{-3}$ Cu-mol$^{-1}$ and 40.1 mJ K$^{-2}$ Cu-mol$^{-1}$ above 1 K, and 19.9 mJ K$^{-3}$ Cu-mol$^{-1}$ and 15.2 mJ K$^{-2}$ Cu-mol$^{-1}$ below 1 K, respectively. The finite $T$-linear coefficients indicate gapless magnetic excitations that have high and low densities of states for the high- and low-temperature phases, respectively. This gapless ground state is transformed to a more ordered state above a magnetic field of approximately 5 T, as evidenced by the gradual vanishment of the $T$-linear term. This critical field is close to $H_{s1} = 4.3$ T for the field-induced transition observed in the first magnetization step.

**Acknowledgements**

This work was partially supported by a Grant-in-Aid in the Priority Area "Novel States of Matter Induced by Frustration (20046010)" from the Ministry of Education, Culture, Sports, Science, and Technology of Japan.

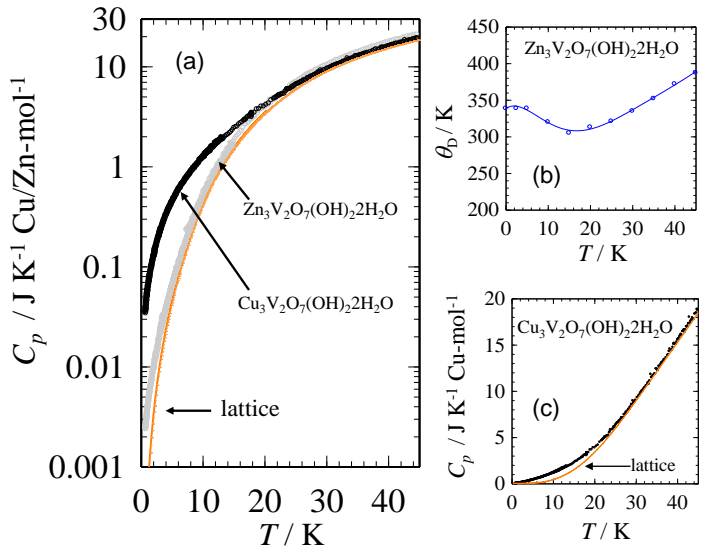

Fig.1

Fig. 1. (a) Temperature dependences of heat capacities of volborthite and the Zn analogue compound shown on a logarithmic scale for $C_p$. (b) Debye temperatures ($\theta_D$) determined at various temperatures below 45 K for $Zn_3V_2O_7(OH)_2 \cdot 2H_2O$. (c) $C_p$ vs $T$ plot for volborthite and a possible lattice component determined on the basis of the data of the Zn compound.

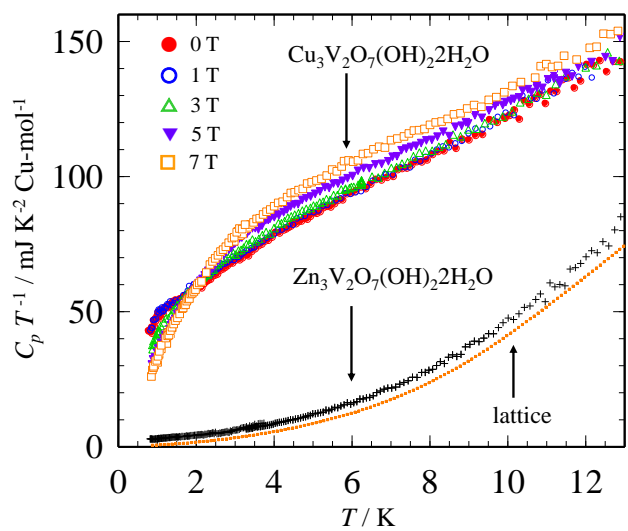

Fig. 2. Temperature dependences of heat capacity divided by temperature for volborthite at magnetic fields between 0 and 7 T.

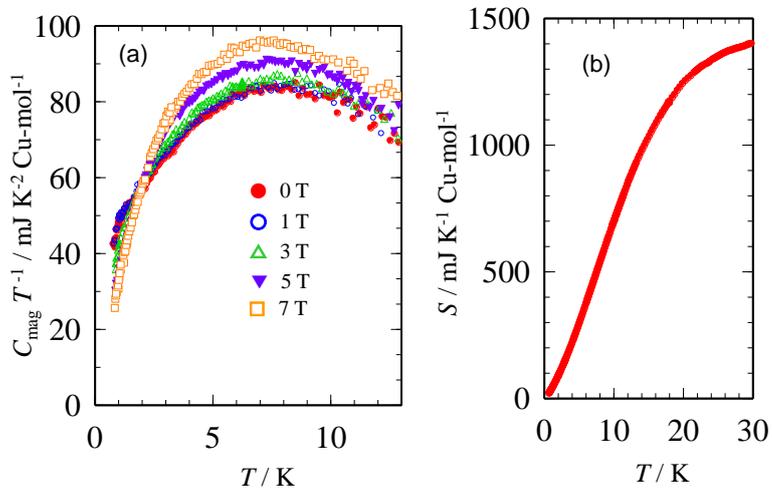

Fig.3

Fig. 3. (a) Temperature dependences of magnetic heat capacity ($C_{mag}$) of volborthite obtained by subtracting the lattice contribution. (b) Temperature dependence of magnetic entropy ($S$) for volborthite under zero magnetic field.

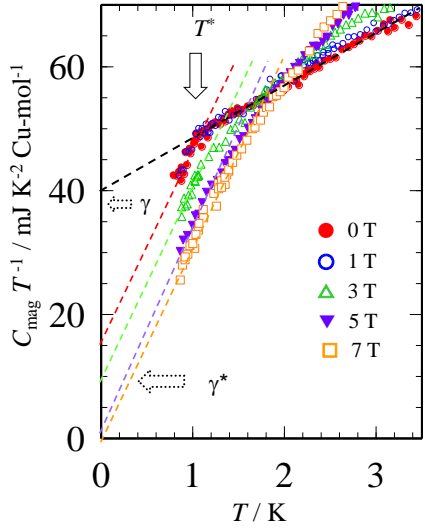

Fig.4

Fig. 4. Magnetic heat capacity of volborthite at low temperature around $T^* = 1$ K in a $C_{mag}T^{-1}$ vs $T$ plot. The extrapolations of the data at $H = 0$ above and below $T^*$ down to $T = 0$ give the coefficients of the $T$-linear term in the magnetic heat capacity, namely, $\gamma$ and $\gamma^*$, respectively. $\gamma^*$ decreases with increasing magnetic field and vanishes above 5 T (see text for details).